\documentclass[
  ,final            
  ,numberedheadings 
  ]
  {aipproc}

\layoutstyle{6x9}

\usepackage{amsmath, amssymb}
\usepackage{pifont}

\begin{document}

\title{Observation and modelling of dusty, low gravity L, and M dwarfs}

\classification{95.30.Wi, 97.10.Ex, 97.20.Vs}
\keywords      {stars: atmospheres, stars: low-mass, brown dwarfs, infrared: stars  }

\author{Andreas Seifahrt}{
 address={Institut f\"ur Astrophysik, Georg-August-Universit\"at, D-37077 G\"ottingen, Germany}
}

\author{Christiane Helling}{
  address={SUPA, School of Phys. and Astron., Univ. of St Andrews, North Haugh, St Andrews,  KY16 9SS, UK}
}

\author{Adam J. Burgasser}{
address={Kavli Institute for Astrophysics and Space Research, Massachusetts Institute of Technology, Building 37, Room 664B, 77 Massachusetts Avenue, Cambridge, MA 02139}
}


\author{Katelyn N. Allers}{
address={Institute for Astronomy, University of Hawai{'}i, 2680 Woodlawn Drive, Honolulu, HI 96822}
}

\author{Kelle L. Cruz}{
address={Department of Astronomy, MS 105-24, California Institute of Technology, Pasadena, CA 91125}
}

\author{Michael C. Cushing}{
address={Institute for Astronomy, University of Hawai{'}i, 2680 Woodlawn Drive, Honolulu, HI 96822}
}


\author{Ulrike Heiter}{
address={Department of Physics and Astronomy, Uppsala University, Box 515, 751 20 Uppsala, Sweden}
}

\author{Dagny L. Looper}{
address={Institute for Astronomy, University of Hawai{'}i, 2680 Woodlawn Drive, Honolulu, HI 96822}
}


\author{S\"oren Witte}{
address={Hamburger Sternwarte, Gojenbergsweg 112, 21029 Hamburg, Germany}
}

\begin{abstract}
Observational facilities allow now the detection of optical and IR
spectra of young M- and L-dwarfs. This enables empirical comparisons
with old M- and L- dwarfs, and detailed studies in comparison with
synthetic spectra. While classical stellar atmosphere physics seems
perfectly appropriate for old M-dwarfs, more physical and chemical
processes, cloud formation in particular, needs to be modelled in the
substellar regime to allow a detailed spectral interpretation.

Not much is known so far about the details of the inset of cloud
formation at the spectral transition region between M and L
dwarfs. Furthermore there is observational evidence for diversity in
the dust properties of objects having the same spectral type. Do we
understand these differences? The question is also how young M- and
L-dwarfs need to be classified, which stellar parameter do they have
and whether degenerations in the stellar parameter space due to the
changing atmosphere physics are present, like in the L-T transition
region.

The Splinter was driven by these questions which we will use to
encourage interactions between observation and theory. Given the
recent advances, both in observations and spectral modelling, an
intensive discussion between observers and theoreticians will create
new synergies in our field.

\end{abstract}

\maketitle


\section{Introduction}
The late-type M and L dwarfs span a critical regime over which both the
internal and atmospheric properties of low mass stellar objects make
important transitions.  Internally, this regime encompasses the hydrogen
burning mass limit ($\sim$ 0.072~M$_{sun}$ for solar metallicity) for ages
typical of the Galactic disk. Young M and L dwarfs in star forming regions
($\tau \sim$ 1-10\,Myr) or young associations ($\tau \sim$ 10-100\,Myr)  are
likely to be among the lowest-mass brown dwarfs currently known, while
their counterparts in the field may
be among the longest-lived low-mass stars in the Galaxy.  The age and mass
of a brown dwarf is directly related to its surface gravity, which can be
discerned in that source's spectral properties.  The atmospheres of M and
L dwarfs are also host to condensable species of refractory elements, which
have a prominent effect on atmospheric opacity and overall spectral energy
distributions.  These condensable constituents of the atmospheric gas are 
likely to form grain particles whose chemical and structural composition and 
spatial distribution are only beginning to be explored in detail
\cite{2001ApJ...556..872A, 2006AA...455..325H}.

Observationally, the near- and mid-infrared spectra and colours of
``normal'' late-type M and L dwarfs are seen to exhibit a wide range of
variations within a given spectral subclass.  $J-K_s$ colours for L dwarfs
with similar optical spectra can vary by over one magnitude, and
near-infrared spectral slopes for these sources exhibit a comparable range
of variation (e.g., \cite{2000AJ....120..447K, 2001ApJ...561L.115M,
2004AJ....127.3553K}).  Such variations have been attributed to changing
cloud characteristics like cloud thickness, inhomogeneous cloud distributions, 
and/or changing opacity properties of the clouds and have stymied efforts 
to derive a complete classification scheme for L dwarfs in the near-infrared (e.g.,
\cite{2002ApJ...564..466G, 2003IAUS..211..355S}).  Specific peculiar features,
such as the presence of enhanced VO and H$_2$O absorption, unusually weak
alkali lines, and peculiar spectral peaks in the near-infrared are
discerned in the spectra of very low-mass M and L dwarfs in young
associations (e.g., \cite{2003ApJ...593.1074G, 2007ApJ...657..511A}) and in some
seemingly ``isolated'' sources (e.g., \cite{2003AJ....126.2421C,
2006ApJ...639.1120K}); these have been attributed to surface gravity
effects.  Even metallicity can play a significant role in shaping the
observed spectra of M and L dwarfs (see contributions in the ``Ultracool
Subdwarfs'' splinter session).   More often than not, M and L dwarf
spectra and colours are a function of several parameters--i.e., effective
temperature, surface gravity, cloud properties, metallicity and possibly
unresolved multiplicity. Disentangling these parameters is essential if
we are to both accurately characterise individual sources of interest and
provide better observational constraints on rapidly advancing theoretical
models.
The contributions below provide a snapshot of the current state-of-the-art
in the characterisation of dusty and low surface gravity M and L dwarfs
from both observational and theoretical perspectives.

\section{Using Near-IR Spectroscopy to Determine the Ages of Young Brown Dwarfs (Allers et al.) }
The near-infrared (near-IR) is the ideal wavelength range for detailed studies of M and L 
type brown dwarfs, whose spectral energy distributions (SEDs) peak in the near-IR, and may be affected by 
attenuation from interstellar dust. Additionally, laser guide star adaptive optics (AO) has 
recently made several remarkable discoveries, many of which can only be followed 
up in the near-IR, which further motivates detailed study of brown dwarfs in the 
near-IR.

Ages (and masses) for young brown dwarfs are typically determined by placing them 
on an H-R diagram using luminosities measured photometrically and effective 
temperatures determined from spectral types and overlaying evolutionary models 
(e.g. \cite{2008arXiv0806.2818P, 2007ApJS..173..104L, 2006AJ....131.3016S}). 
H-R diagram inferred ages rely heavily on the accuracy of evolutionary models 
(which are very uncertain at low masses and young ages), and can be affected by 
binarity, distance uncertainties, dust extinction, occultation from a circumstellar 
disk, and accretion history (see poster by Gallardo et al.). We can avoid these 
problems by using features in the near-IR to determine the ages of our sources.

We have assembled a sample of over 80 moderate resolution (R=750-2000) near-IR spectra 
of brown dwarfs. Our sample includes objects in the 1--3\,Myr old star forming regions 
of Ophiuchus, Lupus, Taurus, ChamI, ChamII, and IC348, the 5\,Myr old Upper Scorpius 
OB Association (includes objects from \cite{2008MNRAS.383.1385L}), the 10 Myr old TW Hydra moving 
group, and the $\sim$Gyr old field population (from \cite{2005ApJ...623.1115C}). Our objects have 
spectral types from M5 to L2. We place our young cluster objects (association ages of 
1--3 Myr) on the H-R diagram using luminosities and effective temperatures from the 
literature, and find our objects have a spread in H-R diagram ages of <1 to over 100\,Myr!
Is the age spread real? Are clusters coeval? Do brown dwarfs have a disparate formation 
history from stars?

To find the ages of our the objects in our sample, we look at the depths of the alkali 
(NaI and KI) lines, which are known to increase with age (e.g. \cite{2007MNRAS.381.1077R, 2006ApJ...639.1120K}). 
We find that the line equivalent widths of our young (1--3\,Myr) 
cluster objects are significantly lower than seen for objects of the same spectral type in 
5\,Myr old Upper Sco or 10 Myr old TW Hydra, this argues that the H-R diagram inferred age 
spread is not real. Using near-IR NaI and KI lines, differences in age of $\sim$3\,Myr can be 
determined for young brown dwarfs.

\section{The {\sc DRIFT-PHOENIX} model grid (Witte et al.)}
  Condensation becomes a major issue in late type dwarfs. In order to
  approximate this influence, the general-purpose stellar
  atmosphere code {\sc PHOENIX} \cite{1999JCoAM.109...41H}
  incorporated the very basic {\sc DUSTY} / {\sc COND} dust model
  \cite{2001ApJ...556..357A}.
  Although these models have been able to improve the synthetic spectra
  and are fairly accurate for $T_\mathrm{eff}$ above 2500\,K ({\sc
  DUSTY}) and $T_\mathrm{eff} <$ 1000\,K  ({\sc COND}), they either
  over- or underestimate the dust cloud in the model
  atmospheres. A more detailed dust treatment is required
  to reproduce observations for effective temperatures between 1000\,K
  and 2500\,K. Therefore, the latest {\sc PHOENIX} version includes the
  phase non-equilibrium {\sc DRIFT} model by \cite{2008A&A...485..547H}, 
  which takes into account nucleation with a subsequent
  kinetic growth and evaporation of  dust grains, made of a mixture of
  7 different solid species, accompanied by gravitational settling and
  element replenishment by convective overshooting \cite{2006AA...455..325H, 2003A&A...399..297W, 2004A&A...414..335W}. 
  The opacities of the mixed dust grains are calculated by
  effective medium and Mie theory \cite{2007IAUS..239..227D}.

  A {\sc DRIFT-PHOENIX} model grid for $T_\mathrm{eff}$=1500...3000\,K,
  ranging over $\log{g}$=3.0...6.0 and [M/H]=$-6.0$...$+0.5$ is almost
  complete. We find a dust cloud structure of five characteristic
  regions in all our models. Starting from the highest altitudes,
  there is a (1) nucleation dominated region, followed by a first
  growth region (2), a region dominated by gravitational settling (3),
  caused by the strong gas phase depletion, a second growth region (4)
  due to the end of nucleation and an evaporation region (5). For
  decreasing effective temperature, the dust clouds become more
  expanded and more dense, resulting in stronger dust features in the
  atmosphere structure and the corresponding spectra. While clouds
  exist up to $T_\mathrm{eff}$=2500K for $\log{g}$=3.0, they persist to
  $T_\mathrm{eff}$=2800\,K for $\log{g}$=6.0, because the clouds are
  shifted to higher gas densities, resulting in a more efficient dust growth.

\section{Young, Low Surface-Gravity L Dwarfs Identified in the Field: A 
Tentative Low-Gravity Spectral Sequence (Cruz et al.)}
We present an analysis of existing optical spectroscopy of 23 L dwarfs that 
display unusual spectral features, including weak FeH molecular absorption 
and weak NaI and KI doublets. All of these features are 
attributable to low-gravity and indicate that these objects are young, 
low-mass brown dwarfs. Twenty-one of these L dwarfs were uncovered during our 
search for nearby, late-type objects using the Two Micron All-Sky Survey while two
were identified in the literature. These spectra form an optical low-gravity 
spectral sequence extending from L0 to L5. Many of these low-gravity L dwarfs 
have southerly declinations and distance estimates within 60\,pc. Their implied youth, 
on-sky distribution, and distances suggest that they are members of nearby, 
intermediate-age ($\sim$10--100\,Myr), loose associations such as the $\beta$~Pictoris 
moving group, the Tucana/Horologium association, and the AB Doradus moving group.
However, before ages and masses can be confidently adopted for any of these 
low-gravity L dwarfs, additional kinematic observations are needed to confirm 
cluster membership.

\section{Crosstalk of dust properties and low-gravity features (Looper et al.)}

The spectroscopic characteristics of young, late M and early L dwarfs have been empirically 
noted in young clusters and nearby moving groups. These features include weak hydrides and 
alkali lines, strong H$_2$O and VO absorption, and a markedly triangular $H$-band in comparison
to normal field dwarfs. The identification of these features has allowed brown dwarfs 
discovered in the field without any known association to be classified as young (on the 
order of less than 100\,Myr). To date, no mid-to-late L dwarfs in young clusters are known, 
leaving it unclear as to whether all or some of these characteristic features of youth persist 
at lower temperatures. Several late L dwarfs, such as 2MASS 2148+4003, have been discovered 
which mimic some signs of youth but not others - a triangular $H$-band and weaker alkali lines 
but having weaker H$_2$O and typical hydride absorption compared to field dwarfs. VO absorption 
no longer persists down to these lower temperatures. However, the interpretation of these 
features as the result of youth is counteracted by their kinematics, which show high proper 
motions and slow rotational velocities, suggesting that these objects are actually old. These 
results show that at late L spectral types these empirical trends should be used with caution 
to classify objects as young in the field. They also highlight the need for identification of 
late L fiducials in clusters.

\section{Atmospheric Parameters of Field L and T Dwarfs (Cushing et al.)}

We have compared the 0.9 to 14.5 micron spectra of 7 L and 2 T dwarfs to
the synthetic spectra generated by the model atmospheres of Marley \&
Saumon \cite{2008ApJ...678.1372C}.  This is the first time L and T dwarf
spectra with such broad wavelength coverage have been compared to
synthetic spectra.  The grid of spectra, computed in chemical
equilibrium, cover from 700 to 2400\,K in steps of 100\,K, have 3
gravities ($\log{g}$ = 4.5, 5.0, 5.5 [cm\,s$^{-2}$]), and 5 sedimention efficiencies
($f_\mathrm{sed}$ = 1, 2, 3, 4, nc).

Overall, the models fit the data well, although there are are
discrepancies near 3 microns in the late L and early T dwarfs (see
however the contribution by D. Stephens). The derived effective
temperatures agree with those derived using evolutionary models and
observed bolometric luminosities \cite{2004AJ....127.3516G}.  Fits to
individual photometric bands almost always produce excellent fits to the
data, but the derived effective temperatures can show a large scatter
compared to those derived by fitting the full spectra; deviations are
typically $\sim$200 K, but can be much larger. In some cases, the resulting
best fitting models are completely inconsistent with the rest of
spectral energy distributions which suggests that atmospheric parameters
derived over narrow wavelength ranges should be considered with caution.
The best fitting model of the very red ($J-Ks=2.05$) L4.5 dwarf 2MASS
J2224-0158 implies that it has very thick condensate clouds (fsed=1) and
a low surface gravity ($\log{g}=4.5$).  However the model does not match the
data well indicating that deriving atmospheric parameters for dusty
and/or low $\log{g}$ L dwarfs using broad wavelength spectra remains
difficult.

\section{Comparative study of cloud formation in brown dwarf atmosphere models
(Helling et al.)}\label{s:cch}
Model simulations for Brown Dwarfs have been challenged by the need of
including clouds which act as opacity source and element sink. Two
main streams developed over the recent years: One approach treats the
cloud in their final state of condensation, namely in
phase-equilibrium (Tsuji, Cooper et al., Allard et al., Ackerman \&
Marley). The other approach treats the actual formation process, hence
treating the dust formation as kinetic process (Helling
\& Woitke). We (\cite{hel2008}) compare our dust cloud models in {\it test case
1} for a given
$(T, p, v_{\rm conv})$ structure excluding all uncertainties in the
radiative transfer treatment, and in {\it test case 2} for given
(T$_{\rm eff}$, $\log$g) combination taking into account the entire
atmosphere simulation. {\it Test case 1} demonstrates that differences
are apparent in our results e.g. in the amount of dust produced, or
the grain size distributions inside the cloud. We further studied the
abundance of certain molecules in the remaining gas phase, and the
phase-non-equilibrium approach would always produce the highest
abundances concerning those element involved into dust formation in
and above the cloud layer.  {\it Test case 2} allows to compare
e.g. integrated fluxes where we find that single models might suggest
extreme values but the mean values over all models do recover the
spectral type of the object for our sample of four models for a given
stellar parameter combination. Note that e.g. {\sc Settl-Phoenix}
would not reproduce observations for L dwarfs with the code-version
used in this comparison study.

\section{Observations and analysis of low-mass benchmark stars (Heiter et al.)}

We report on a project to establish a set of benchmark stars in the M dwarf 
region. The goals of this project are two-fold: 1) We aim to improve the
effective temperature scale for M dwarfs. 2) We seek to establish a reliable 
calibration of metallicity, in particular at the high-metallicity end.
Our means to achieve these goals are based on high-resolution spectroscopic 
observations in the red and near-infrared (with UVES and CRIRES at ESO's
VLT). The data are analysed using synthetic spectra based on MARCS stellar 
atmosphere models \cite{2008A&A...486..951G} and recent atomic and molecular line lists in order to
constrain element abundances and temperatures of the program stars. The targets 
span spectral classes from M0 to M4 and fall in the metallicity range of $-0.5$ 
to $+0.5$ dex. A significant fraction of the program stars are in binary systems 
with companions of earlier type, and both components are being studied in order 
to better constrain the stellar parameters. The results of this study will be 
applied in two different areas: 1) The exploration of the metallicity 
distribution for M dwarf planet hosts in comparison to solar-type hosts will be 
made possible \cite{2003AJ....126.2015H}. 2) For the analysis of data obtained by ESA's Gaia mission, M dwarf 
benchmark stars are needed to calibrate the astrophysical parameter determination 
and to provide templates for radial velocity measurements. Although we confine 
ourselves in the first phase of this project to early M types where dust formation 
in the atmospheres is negligible, an extension of the program stars towards lower 
masses is anticipated in later phases.

\section{Conclusions \& Outlook}

The work presented at this splinter demonstrate the remarkable progress on the field
of young low mass stars and brown dwarfs in the last years, both from new observations 
and from sophisticated theory. 
So far, the determination of the physical properties of young low mass objects 
relied fully on theoretical predictions of their colours and luminosities. But 
theory remained largely unverified due to the scarcity of young calibrators. Decent 
spectra existed only for a handful of these young brown dwarfs and allowed only 
a first assessment of their spectral features to determine their properties and 
to identify their age. Now we see that observers provide us with an ever growing 
number of spectra of young brown dwarfs, allowing for the first time a systematic 
study of their observational properties. 

With plurality comes diversity and we are faced with the challenge of developing a 
multi-dimensional classification scheme that includes at least three properties, 
i.e. surface gravity, metallicity, and effective temperature. The fact that all 
these properties are interlinked and a function of age adds to the level of complexity 
involved in this task. The number of ``peculiar'' objects is growing, indicating that 
we have either not fully understood the possible crosstalk of different properties or 
that brown dwarfs are showing more signs of individualism than stars. This should
not dilute the need for the in-depth characterisation of a few benchmark objects.
Especially the determination of accurate metallicities for these objects will be
difficult and provides already a challenge for regular (old) M dwarfs.

On the other hand, the theoretical description of the spectral features of young and 
dusty objects has made tremendous progress. Different modeller groups were present showing 
their results and demonstrated the diversity in the field. Models with different physical 
assumptions on dust treatment, namely assuming phase-equilibrium (Marley et al., Freytag et al.) 
or applying  kinetic treatments (Helling et al., Witte et al.), produce very different 
predictions on the spectral properties of brown dwarfs. Observers should feel encouraged 
to not only hook with one particular model but compare different ``flavours'' and also keep 
in mind the individual ``trust range'' of the models when assigning physical properties, such 
as mass and age, based on spectral fits. Note that the comparison study included here concerns  
atmosphere simulations only, and not the modeling of the Brown Dwarfs evolution.

Although the drive in our field goes into ever cooler objects down to the planetary regime, 
young brown dwarfs keep providing surprises. Key issues still remain little understood such 
as the inset of dust formation at the M/L type transition and its effect on the observable 
properties of these objects. 
New telescopes and instruments, suitable for studying brown dwarfs, become available in the 
next decade, such as WISE and NIRCam at the \textit{JWST} or instruments at the next generation 
of ground based large telescopes, such as the ELT. Multi-wavelength studies in the near- and 
thermal-Infrared with unprecedented sensitivities and/or extremely high spatial and spectral 
resolution will provide further and new insides into the atmosphere chemistry of such ultra-cool 
objects but also into the formation scenario of brown dwarfs.


\begin{theacknowledgments}
The authors wish to thank the organisers of Cool Stars 15 for hosting this splinter session,
as well as Mark Marley, Mark McCaughrean and Bernd Freytag for their contributions to the
splinter. AS acknowledges financial support from the Deutsche Forschungsgemeinschaft under 
DFG RE 1664/4-1. UH acknowledges financial support from the Swedish National Space Board.
\end{theacknowledgments}



\bibliographystyle{aipproc}   


\begin{thebibliography}{10}
\providecommand{\enquote}[1]{``#1''}
\expandafter\ifx\csname url\endcsname\relax
  \def\url#1{\texttt{#1}}\fi
\expandafter\ifx\csname urlprefix\endcsname\relax\def\urlprefix{URL }\fi


\bibitem{2001ApJ...556..872A}
A.~S. {Ackerman}, and M.~S. {Marley} \ 2001, \emph{ApJ}, \textbf{556}, 872--884

\bibitem{2006AA...455..325H}
Ch.~{Helling}, and P.~{Woitke} \ 2006, \emph{A\&A} , \textbf{455}, 325--338

\bibitem{2000AJ....120..447K}
J.~D. {Kirkpatrick}, I.~N. {Reid}, J.~{Liebert}, J.~E. {Gizis}, A.~J.
  {Burgasser}, D.~G. {Monet}, C.~C. {Dahn}, B.~{Nelson}, and R.~J. {Williams}\ 
  2000, \emph{AJ} \textbf{120}, 447--472

\bibitem{2001ApJ...561L.115M}
I.~S. {McLean}, L.~{Prato}, S.~S. {Kim}, M.~K. {Wilcox}, J.~D. {Kirkpatrick},
  and A.~{Burgasser} \ 2001, \emph{ApJL} \textbf{561}, L115--L118

\bibitem{2004AJ....127.3553K}
G.~R. {Knapp}, S.~K. {Leggett}, X.~{Fan}, M.~S. {Marley}, T.~R. {Geballe},
  D.~A. {Golimowski}, D.~{Finkbeiner}, J.~E. {Gunn}, J.~{Hennawi},
  Z.~{Ivezi{\'c}}, R.~H. {Lupton}, D.~J. {Schlegel}, M.~A. {Strauss}, Z.~I.
  {Tsvetanov}, K.~{Chiu}, E.~A. {Hoversten}, K.~{Glazebrook}, W.~{Zheng},
  M.~{Hendrickson}, C.~C. {Williams}, A.~{Uomoto}, F.~J. {Vrba}, A.~A.
  {Henden}, C.~B. {Luginbuhl}, H.~H. {Guetter}, J.~A. {Munn}, B.~{Canzian},
  D.~P. {Schneider}, and J.~{Brinkmann} \ 2004, \emph{AJ} \textbf{127}, 3553--3578

\bibitem{2002ApJ...564..466G}
T.~R. {Geballe}, G.~R. {Knapp}, S.~K. {Leggett}, X.~{Fan}, D.~A. {Golimowski},
  S.~{Anderson}, J.~{Brinkmann}, I.~{Csabai}, J.~E. {Gunn}, S.~L. {Hawley},
  G.~{Hennessy}, T.~J. {Henry}, G.~J. {Hill}, R.~B. {Hindsley}, {\v
  Z}.~{Ivezi{\'c}}, R.~H. {Lupton}, A.~{McDaniel}, J.~A. {Munn}, V.~K.
  {Narayanan}, E.~{Peng}, J.~R. {Pier}, C.~M. {Rockosi}, D.~P. {Schneider},
  J.~A. {Smith}, M.~A. {Strauss}, Z.~I. {Tsvetanov}, A.~{Uomoto}, D.~G. {York},
  and W.~{Zheng} \ 2002, \emph{ApJ} \textbf{564}, 466--481

\bibitem{2003IAUS..211..355S}
D.~C. {Stephens}, \enquote{The Classification of L Dwarfs: Is It Based on
  Clouds or Temperature?} in {Brown Dwarfs}, edited by
  E.~{Mart{\'{\i}}n}\ 2003, vol. 211 of \emph{IAU Symposium}, p. 355.

\bibitem{2003ApJ...593.1074G}
 N.~I. {Gorlova}, M.~R. {Meyer}, G.~H. {Rieke}, and J.~{Liebert}\ 2003, \emph{ApJ}
  \textbf{593}, 1074--1092

\bibitem{2007ApJ...657..511A}
K.~N. {Allers}, D.~T. {Jaffe}, K.~L. {Luhman}, M.~C. {Liu}, J.~C. {Wilson},
  M.~F. {Skrutskie}, M.~{Nelson}, D.~E. {Peterson}, J.~D. {Smith}, and M.~C.
  {Cushing} \ 2007, \emph{ApJ} \textbf{657}, 511--520

\bibitem{2003AJ....126.2421C}
K.~L. {Cruz}, I.~N. {Reid}, J.~{Liebert}, J.~D. {Kirkpatrick}, and P.~J.
  {Lowrance} \ 2003, \emph{AJ} \textbf{126}, 2421--2448

\bibitem{2006ApJ...639.1120K}
J.~D. {Kirkpatrick}, T.~S. {Barman}, A.~J. {Burgasser}, M.~R. {McGovern}, I.~S.
  {McLean}, C.~G. {Tinney}, and P.~J. {Lowrance} \ 2006, \emph{ApJ} \textbf{639},
  1120--1128


\bibitem[Peterson et al.(2008)]{2008arXiv0806.2818P} D.~E. {Peterson}, S.~T. {Megeath}, K.~L. {Luhman}, J.~L. {Pipher}, J.~R. {Stauffer}, D. {Barrado y Navascues}, J.~C. {Wilson}, M.~F. {Skrutskie}, M.~J. {Nelson}, and J.~D. {Smith} \ 2008, ArXiv e-prints, 806, arXiv:0806.2818 

\bibitem[Luhman(2007)]{2007ApJS..173..104L} K.~L. Luhman\ 2007, \emph{ApJS}, 
\textbf{173}, 104 

\bibitem[Slesnick et al.(2006)]{2006AJ....131.3016S} C.~L. Slesnick, J.~M. Carpenter, and L.~A. Hillenbrand\ 2006, \emph{AJ}, \textbf{131}, 3016 

\bibitem[Lodieu et al.(2008)]{2008MNRAS.383.1385L} N. Lodieu, N.~C. Hambly, R.~F. Jameson, and S.~T. Hodgkin\ 2008, \emph{MNRAS}, \textbf{383}, 1385 

\bibitem[Cushing et al.(2005)]{2005ApJ...623.1115C} M.~C. Cushing, J.~T. Rayner, and W.~D. Vacca\ 2005, \emph{ApJ}, \textbf{623}, 1115

\bibitem[Riddick et al.(2007)]{2007MNRAS.381.1077R} F.~C. Riddick, P.~F. Roche, and P.~W. Lucas\ 2007, \emph{MNRAS}, \textbf{381}, 1077 



\bibitem[Hauschildt 
\& Baron(1999)]{1999JCoAM.109...41H} P.~H. Hauschildt, and E. Baron\ 1999, 
\emph{Journal of Computational and Applied Mathematics}, \textbf{109}, 41 

\bibitem[Allard et al.(2001)]{2001ApJ...556..357A} F. Allard, P.~H. Hauschildt, 
D.~R. Alexander, A. Tamanai, and A. Schweitzer\ 2001, \emph{ApJ}, \textbf{556}, 357

\bibitem[Helling et 
al.(2008)]{2008A&A...485..547H} Ch. Helling, P. Woitke, and W.-F. Thi\ 2008, \emph{A\&A}, \textbf{485}, 547 

\bibitem[Woitke 
\& Helling(2003)]{2003A&A...399..297W} P. Woitke, and Ch. Helling\ 2003, \emph{A\&A}, \textbf{399}, 297

\bibitem[Woitke 
\& Helling(2004)]{2004A&A...414..335W} P. Woitke, and Ch. Helling\ 2004, \emph{A\&A}, \textbf{414}, 335 

\bibitem[Dehn et al.(2007)]{2007IAUS..239..227D} M. Dehn, Ch. Helling, P. Woitke, and P. Hauschildt\ 2007, \emph{IAU Symposium}, \textbf{239}, 227 


\bibitem[Cushing et al.(2008)]{2008ApJ...678.1372C} M.~C. {Cushing}, M.~S. {Marley}, D. {Saumon}, B.~C. {Kelly}, W.~D. {Vacca}, J.~T. {Rayner}, R.~S. {Freedman}, K. {Lodders}, and T.~L. {Roellig}\ 2008, \emph{ApJ}, \textbf{678}, 1372

\bibitem[Golimowski et al.(2004)]{2004AJ....127.3516G} D.~A. {Golimowski}, S.~K. {Leggett}, M.~S. {Marley}, X. {Fan}, 
T.~R. {Geballe}, G.~R. {Knapp}, F.~J. {Vrba}, A.~A. {Henden}, C.~B. {Luginbuhl}, H.~H. {Guetter}, J.~A. {Munn}, B. {Canzian},
W. {Zheng}, Z.~I. {Tsvetanov}, K. {Chiu}, K. {Glazebrook}, E.~A. {Hoversten}, D.~P. {Schneider}, and J. {Brinkmann}\ 2004, \emph{AJ}, \textbf{127}, 3516


\bibitem{hel2008}
Ch. Helling, A. Ackerman, F. Allard, M. Dehn, P. Hauschildt,
D. Homeier, K. Lodders, M. Marley, R. Rietmeijer, T. Tsuji, P. Woitke
2008, \emph{MNRAS}, in press




\bibitem[Gustafsson et 
al.(2008)]{2008A&A...486..951G} B. Gustafsson, B. Edvardsson, K. Eriksson, U.~G. J{\o}rgensen, {\AA} Nordlund, and B. Plez\ 2008, \emph{A\&A}, \textbf{486}, 951

\bibitem[Heiter 
\& Luck(2003)]{2003AJ....126.2015H} U. Heiter, and R.~E. Luck\ 2003, \emph{AJ}, \textbf{126}, 2015 

\end{thebibliography}

\IfFileExists{\jobname.bbl}{}
 {\typeout{} \typeout{******************************************}
  \typeout{** Please run "bibtex \jobname" to optain} \typeout{** the
  bibliography and then re-run LaTeX} \typeout{** twice to fix the
  references!}  \typeout{******************************************}
  \typeout{} }

\end{document}